\documentstyle[12pt,epsfig]{article}
\newcommand\slv{v\kern-5pt\raise1pt\hbox{$\scriptstyle/$}\kern1pt}

\begin{document}
\thispagestyle{empty}
\begin{flushright}
BudkerINP-96-21\\
WUE-ITP-97-029\\
hep-ph/97XXXX\\
July 1997\\
\end{flushright}
\vspace{0.5cm}
\begin{center}
{\Large\bf On the resummation of double logarithms}\\[.3cm]
{\Large\bf in the process $H\to\gamma\gamma$}\\
\vspace{1.3cm}
{\bf M.I. Kotsky$^{a,}$\footnote{kotsky@inp.nsk.su}
and O.I. Yakovlev$^{b,}$\footnote{\offinterlineskip
o\_yakovlev@physik.uni-wuerzburg.de; on leave from {\em a}. }\\[1cm]
{\em a}- Budker Institute of Nuclear Physics (BINP),\\
pr. Lavrenteva 11, Novosibirsk, 630090, Russia\\[0.5cm]
{\center and}\\
{\em b}- Instit\"ut f\"ur Theoretische Physik II,\\[.2cm]
Am Hubland, D-97074, W\"urzburg, Germany  }\\ 
\vspace{1cm}
\vspace{1cm}
\end{center}
\begin{abstract}\noindent
The decay of the Higgs boson into two photons is discussed.
Using Sudakov technique, we study a new 
type of double logarithms 
in the case of the light quark loop. 
 First, we examine the origin of such logarithms at
the one loop level.
Then we perform a two-loop analysis, and finally we 
present an explicit 
result of the resummation of double logarithms in all order in QCD.
The phenomenological applications are discussed.   
\end{abstract}
\vspace{0.5cm}

\newpage

\section{Introduction}
The neutral scalar Higgs boson $H$ is an important ingredient 
of the Standard Model (SM) and is the only SM elementary 
particle which  
has not been detected so far. 
The lower limit on $m_H$ of approximately 70 GeV was obtained 
from direct searches at LEP \cite{Limit1}. The one sided 
$95\%$ confidence level upper limit on $m_H$ is $420$ GeV 
\cite{Limit1}.
The perturbative unitarity suggests an 
upper limit on the Higgs boson mass to be about $1$ TeV 
\cite{Limit2}.

The properties of the Higgs particle were studied in 
details by many authors
( see, for example reviews \cite{Hunter,Dawson,Kniehl,Zerwas1,
Djouadi,Zerwas4,Spira} and references therein).

 The decay of the Higgs boson into two photons will play
an important and unique role.
It is of interest for the following reasons:
\begin{itemize} 
\item for the light Higgs masses this decay mode provides 
a clear signature for the search at hadron colliders  
\item the $\gamma\gamma$ width determines the cross section for 
the Higgs production in $\gamma\gamma$ collisions 
\cite{Ginzburg,Habber,Khoze}
\item it is well known that the $H\to\gamma\gamma$ 
vertex can serve as a counter
of the particles with masses larger than the Higgs boson mass.
\end{itemize}
 It is widely accepted that 
a detailed experimental investigation will be performed 
at the Next Linear ${\bf e^+~e^-}$ Collider which will also operate 
in the photon-photon mode \cite{Ginzburg, Telnov}, as a Photon Linear 
Collider ( see the review of physical applications in \cite{PhysPhot}). 
The possibility of studying scalar Higgs boson 
in the s-channel production, for example in a reaction
\begin{eqnarray}
\gamma + \gamma \to H \to q\bar q   
\end{eqnarray} 
is not so obvious, because of large background from 
direct quark-antiquark 
 production. Fortunately, it was 
demonstrated that using the polarizations of initial photons one can 
suppress background process at high energy despite the QCD corrections 
to the background process is large \cite{Habber,Khoze,Fadin}. 
If the task of efficient suppression will be achieved, one can 
hope that the Higgs will be investigated in detail in $\gamma\gamma$
collisions and the idea of precise determination of the 
$H\gamma\gamma$ vertex will then become an important 
phenomenological issue.

Therefore, the $H\gamma\gamma$ vertex should be 
investigated as thoroughly
as possible in the SM as well in the Supersymmetric extension of the 
SM (SSM).    

In the present paper we study QCD corrections to the decay rate 
in the limit of the small ratio of the quark  and Higgs masses, 
$m_q$ and $m_H$. In this limit, an interesting type of QCD double 
logarithms (DL) appears and we make an attempt of their resummation.
The fact of the emergency of DL in the process $H\to \gamma\gamma$ was
pointed in \cite{YaMe}. The similar phenomena
was found and studied  in \cite{Jikia,Fadin} 
in the another physical process, $\gamma\gamma\to q\bar q$ 
in the $J_Z=0$
channel.
 The authors of \cite{Fadin} have performed a careful analysis 
of the origin of the DL appearance  and studied DL terms 
up to two-loop accuracy. 

There are two reasons why the problem of DL in $H\to 
\gamma\gamma$ seems 
to be worth of being investigated. The first reason is 
more or less academical. 
Once we have found a process where a new type of DL 
(in a sense that it is 
not a Sudakov-like logarithms) appear it is interesting 
to understand their origin and even resum them. 
From QED and QCD we know a few examples 
where DL appear and can be resummed in observable quantities 
(see, for example, reviews \cite{Akh, Mull}, and also 
recent paper \cite{Fadin} and references therein).
We claim now that the $H\to \gamma\gamma$ decay considered here 
is the next new example.     
The second reason of interest 
is phenomenological. As we mentioned above the precise 
measurement of the $H\to \gamma\gamma$ decay width will 
be performed at 
Next Linear Collider.
We note at the very beginning that in the SM the contribution 
of light quark to the amplitude is proportional to the mass 
of this quark 
and therefore is small. 
However with an improvement of precision, 
the contribution of
heaviest from 'light quarks' , the b-quark , can become visible 
in future precise measurements at NLC.
In addition we want to stress that in the SSM the coupling of the 
$b$ quark with the Higgs may be large due to an additional 
large factor $tg(\beta )$ and that the contribution
of the $b$ quark at large $tg(\beta )$ may be as important as 
the contribution of top quark or $W$ boson. 
   
In present paper we start with the one-loop analysis 
and demonstrate that 
one-loop and two-loop contributions  contain a new sort 
of double logarithms
$$\xi=\alpha_s\log^2\left( \frac{m_q^2}{m_H^2} \right). $$ 
If the ratio $m_q/m_H$ 
is quite small the parameter $\xi $ can be large enough 
and the QCD expansion 
in $\xi$ is not valid anymore. For example, considering a situation 
when the Higgs has a mass $m_H=500~GeV$ and decays into two 
photons through $b$-quark loop and $b $ quark has a mass 
$m_b=4.5~GeV$, we see that indeed $\xi \approx 1$ and resummation of 
double logarithms is necessary.   In the present paper 
we show that this 
goal becomes feasible when the physical origin of DL 
is understood and QCD series are properly organized.   
First, we examine an origin of such log's by using Sudakov technique 
and then perform a two-loop calculation in DL approximation.
The resummation of DL terms can be done 
and we present an exact answer in DLA.

The paper is organized as follows. In the section 2 we discuss the
one-loop amplitude and the two-loop QCD correction. 
Section 3 contains the resummation of the DL contributions 
in all orders the perturbative QCD and a brief discussion.

\section{One and two loop results}

We start with considering the amplitude of the process 
$H \to \gamma\gamma$ 
at leading one-loop level.
The amplitude of the Higgs decay into two photons through quark loop 
can be presented in the following form
\begin{equation}\label{1}
M^{(quarks)}(H \rightarrow \gamma\gamma) = 
{e_1^*}_{\mu}{e_2^*}_{\nu}d^{\mu\nu}
\left( G_F \sqrt{2} \right)^{1/2} 
\frac{\alpha}{4\pi}N_c \sum_q Q_q^2 F(x_q),
\end{equation}
where the tensor $d_{\mu\nu}$ is defined as
\begin{equation}\label{2}
d^{\mu\nu} = (k_1k_2)g^{\mu\nu} - k_2^{\mu}k_1^{\nu},
\end{equation}
$e_1,e_2$ and $k_1,k_2$ are polarization vectors and momenta 
of the photons,
$N_c$ - number of colors, $Q_q$ - the electric charge of the 
quark $q$ and
the variable $x_q$ is defined by the following relation:
\begin{equation}\label{3}
x_q = \frac{m_q^2}{m_H^2}.
\end{equation}

The corresponding decay width was calculated in 
\cite{Ellis, Vain} at 
the one-loop level. 
In our notations,  the quark formfactor $F$ defined by 
Eqs. (2, 3) reads
$$
F(z_q) = \frac{1}{z_q}\left( \left( 1 - \frac{1}{z_q} \right) \left[
\ln\left( \frac{\sqrt{z_q} + \sqrt{z_q - 1}}{\sqrt{z_q} - 
\sqrt{z_q - 1}}
\right) - i\pi \right]^2 - 4 \right),\ \ \ z_q = 
\frac{m_H^2}{4m_q^2} =
\frac{1}{4x_q} > 1.
$$
Now we demonstrate how to extract the double logarithmic
contribution from the amplitude using the Sudakov technique.
To understand a basic properties of the amplitude in DLA 
we consider first
a scalar part of amplitude without numerator and then discuss the 
structure of the numerator separately. 
The scalar part of the amplitude reads
\begin{equation}\label{4}
{\tilde F}(x_q) = \frac{-2im_H^2}{\pi^2}\int\limits_{-\infty}^{\infty}
\frac{d^4q}
{\left[ (k_1 + q)^2 - m_q^2 + i0 \right]
\left[ (k_2 - q)^2 - m_q^2 + i0 \right]
\left[ q^2 - m_q^2 + i0 \right]}.
\end{equation} 
This integral is very simple and can be calculated explicitly.
However, we are interested in terms containing double logarithms.
The most suitable and simplest method of calculation was suggested by 
Sudakov in \cite{Sudakov}. 
It consists in a direct integration in the momentum
space splitting the internal momenta into the parallel 
and perpendicular 
components to the plane of the external momenta $k_1$ and $k_2$.
To this end, one introduces new variables $\alpha$ and $\beta$ 
using Sudakov-decomposition for the vector $q$
\begin{equation}\label{7}
q = \alpha k_1 + \beta k_2 + q_{\perp},\ \ \ \ \ d^4q = \frac{m_H^2}{2}
d\alpha d\beta d^2q_{\perp}.
\end{equation}
We obtain for ${\tilde F}$ (see also consideration 
of the similar integral in DLA in Ref. [15])
$$
{\tilde F}(x_q) = \frac{-i}{\pi}
\int\limits_{-\infty}^{\infty}\int\limits_{-\infty}^{\infty}
d\alpha d\beta \int\limits_0^{\infty} d\rho
$$
\begin{equation}\label{8}
\times \frac{1}{\left[ \beta - (x_q + \rho - \alpha\beta ) + i0 \right]
\left[ \alpha + (x_q + \rho - \alpha\beta ) - i0 \right]\left[
(x_q + \rho - \alpha\beta ) - i0 \right]},
\end{equation}
where
\begin{equation}\label{9}
\rho = \frac{-q_{\perp}^2}{m_H^2}.
\end{equation}
In Eq. (\ref{8}) we have performed an integration over the angles 
in the transverse subspace, using the fact that 
the integrand does not contain a
dependence on these variables. It is then evident from (\ref{8}), that
the double logarithmic contribution at $x_q \ll 1$ comes from the 
integration
region where
\begin{equation}\label{10}
\rho, x_q \ll |\alpha|, |\beta| \ll 1.
\end{equation}
To carry out the integration over $\rho$ we may apply the relation
\begin{equation}\label{12}
\frac{1}{\left[ (x_q + \rho - \alpha\beta ) - i0 \right]} =
P\frac{1}{x_q + \rho - \alpha\beta} + i\pi\delta(x_q + \rho -
\alpha\beta).
\end{equation}
In the above expression, $P$ means the principal value. 

One can check that the first term in Eq. (\ref{12}) does not contribute to
the DL asymptotics of the formfactor and we do not consider it here.
 We get for the ${\tilde F}$ in DLA
\begin{equation}\label{13}
{\tilde F}(x_q) = \int\limits_{-\infty}^{\infty} 
\frac{d\alpha}{\alpha}\frac{d\beta}{\beta}d\rho\theta(
\rho) \delta\left( \rho - (\alpha\beta - x_q) \right) = 
\int\limits_{-1}^1
\int\limits_{-1}^{1} \frac{d\alpha}{\alpha} 
\frac{d\beta}{\beta} \theta(\alpha\beta -
x_q).
\end{equation}
In the last relation, we replace the strong inequality (\ref{10}) 
 by usual inequality, limiting the integration region 
in accordance with
the logarithmic character of integration.

Let us now consider the numerator of the amplitude. 
It is clear from presented derivation
that only the soft momenta $q$ give contribution to DL.
That means that we may neglect the dependence on $q$ 
in the numerator of
the integrand:   
\begin{equation}\label{14}
\frac{4x_q}{2m_H^4m_q} d_{\mu\nu} Sp \left[ \gamma^{\mu}({\hat k_1} 
+ {\hat q}
+ m_q)({\hat k_2} - {\hat q} - m_q)\gamma^{\nu}({\hat q} + m_q) 
\right] 
= 4x_q.
\end{equation}
We see that the numerator is proportional to the second power 
of the small quark mass. One power of the mass appears because of the 
Yukawa coupling constant and 
the second power is due to the trace. 
It is important here that only the scalar term of the numerator of the
$t$-channel quark propagator
$({\hat q} + m_q)/(q^2 - m^2_q)$ 
contributes to the amplitude.
That has a simple physical explanation. 
The Higgs particle has spin zero and positive parity. 
Therefore, the initial and final total momenta of the
subprocess $\bar q q \to \gamma \gamma $ are zero. 
And, the amplitude is  suppressed by the mass of the light quark 
$(O(m_q/E_q))$ 
at high energy due to the nonconservation of quark helicity.
We see that the quantum numbers of the external particles
force to have a situation when the $t-$channel quark propagator 
behaves like a scalar boson propagator, which is singular enough 
to produce DL.
Really , the scalar part of the quark propagator 
is more singular then the vector part in the limit of the soft 
momenta and this fact 
guarantees the appearance of the new non-Sudakov DL. 
The possibility of the emergency of such DL in the mass suppressed
amplitudes due to soft quark have been pointed and 
explained in \cite{Fadin},
where helicity nonconserving amplitude $\gamma\gamma 
\rightarrow q\bar q$ was studied in DLA.

Despite the same origin the observed double logarithms
in the $H\to \gamma\gamma$ process  demonstrate some new 
peculiarities. 
For example, this phenomena 1) occurs not in the differential 
distribution or in the special helicity amplitude but rather in 
the decay width, 2) it occurs in the quark loop, 
but not as a correction  to the Born amplitude. 
Therefore, the discussed phenomena is new in a sense that it  
appears in a very specific and unique circumstances, and of course, 
represents a new manifestation of the non-Sudakov type of DL. 

The integration in Eq. (\ref{13}) is trivial. Taking into account
expression (\ref{14}) for the numerator we get the formfactor $F(x_q)$
in the DLA:
\begin{equation}\label{14a}
F(x_q) = 4x_q\ln^2(x_q).
\end{equation}

Now we consider a QCD correction 
to the formfactor $F$ defined by
Eq. (\ref{1}). This QCD correction was calculated in
Refs. \cite{Zerwas3,YaMe,Japan,Zerwas2}. In the limit of small 
quark masses 
the correction has a simple form
\begin{equation}\label{16}
F(x_q) = F(x_q)^{(one-loop)}\left( 1 - \frac{1}{6}
\frac{\alpha_sC_F}{4\pi}
\ln^2(x_q) + ... \right),
\end{equation}
where $C_F$ is given by relation
\begin{equation}\label{17}
t^at^a = C_FI,
\end{equation}
and $t^a$ is the color SU(N) group generators in the fundamental
representation ($C_F = 4/3$ for QCD).
\begin{figure}
\begin{center}
\epsfig{figure=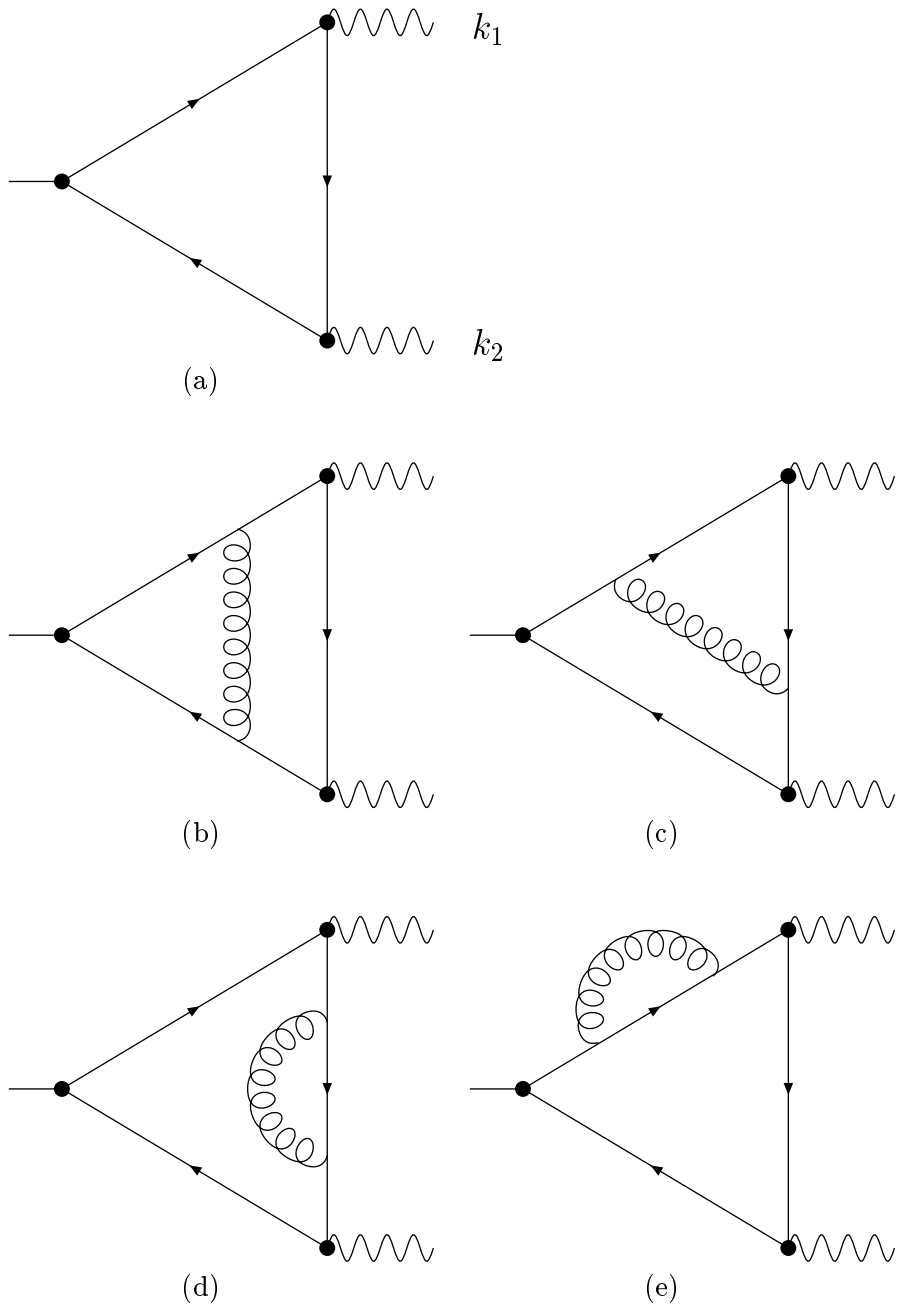}
\end{center}
\caption{Feynman diagrams corresponded to the amplitude
(\ref{1}): (a) one-loop diagram corresponded to the zeroth order in 
$\alpha_s$, (b-e) two-loop diagrams corresponded to the first order 
in $\alpha_s$. External wavy lines represent photons, internal wavy 
lines - gluons, internal solid lines - quarks, external solid lines are
the Higgs bosons.}
\end{figure}
There are four different topologies of the two-loop diagrams, 
which are shown in Fig. 1 (b-e). 
Here we calculate them in the DL approximation by using Sudakov
technique. 

Let us start from diagram 1b and consider first a scalar integral.
Here and below we always use the Feynman gauge for the gluon fields;
 the 
gluon propagator in this gauge is
\begin{equation}\label{17a}
\frac{-i\delta_{ab}g^{\mu\nu}}{k^2 + i0}.
\end{equation}
The scalar part of the $F^{(1b)}(x_q)$ reads
$$
{\tilde F}^{(1b)}(x_q) = \left( \frac{-2im_H^2}{\pi^2} \right)^2 
\int\limits_{-\infty}^{\infty} \frac
{d^4q d^4Q}{\left[ (k_1 + q)^2 - m_q^2 + i0 \right]
\left[ (k_1 + Q)^2 - m_q^2 + i0 \right]}\times
$$
\begin{equation}
\frac{1}{\left[ (k_2 - q)^2 - m_q^2 + i0 \right]
\left[ (k_2 - Q)^2 - m_q^2 + i0 \right]
\left[ q^2 - m_q^2 + i0 \right]\left[ (Q - q)^2 - m_q^2 + i0 \right]}.
\end{equation}
In the Sudakov variables
\begin{eqnarray}
q = \alpha k_1 + \beta k_2 + q_{\perp} \quad \mbox{and} \quad
Q = A k_1 + B k_2 + Q_{\perp},
\end{eqnarray}
we have
$$
{\tilde F}^{(1b)}(x_q) = - \int\limits_{-\infty}^{\infty} 
\frac{d\alpha d\beta d^2q}{\pi^2}
\frac{1}{\left[ \beta - (x_q + {\vec q}^2 - \alpha \beta) + i0 \right]
\left[ \alpha + (x_q + {\vec q}^2 - \alpha \beta) - i0 \right]}\times
$$
$$
\frac{1}{\left[ (x_q + {\vec q}^2 - \alpha \beta) - i0 \right]}
\int\limits_{-\infty}^{\infty}
\frac{dAdBd^2Q}{\pi^2}\frac{1}{\left[ B - (x_q + {\vec Q}^2 - AB) + i0
\right]}\times
$$
\begin{equation}\label{18a}
\frac{1}{\left[ A + (x_q + {\vec Q}^2 - AB) - i0 \right]\left[ \left(
({\vec Q} - {\vec q})^2 - (A - \alpha)
(B - \beta) \right) - i0 \right]},
\end{equation}
where we have switched to dimensionless Euclidian vectors ${\vec Q}$ 
and
${\vec q}$, noticing that the transverse momenta are spacelike. Note 
that
the external integral in Eq. (\ref{18a}) coincides with the one loop 
one.
It is then evident that in order to get the DL contribution 
one should integrate over the following
region
$$
{\vec q}^2, x_q \ll |\alpha|, |\beta| \ll 1,\ \ \ {\vec q}^2 \ll
 {\vec Q}^2,
$$
\begin{equation}\label{18b}
{\vec Q}^2, |\alpha| \ll |A| \ll 1,\ \ \ {\vec Q}^2, |\beta| 
\ll |B| \ll 1.
\end{equation}
Making corresponding simplifications in Eq. (\ref{18a}), performing
integration over ${\vec Q}$ and ${\vec q}$ with the help 
of Eq. (\ref{12}) 
and  taking into account only the absorptive 
parts of the corresponding propagators, we get
\begin{equation}\label{18c}
{\tilde F}^{(1b)}(x_q) = \int\limits_{-1}^1 \frac{d\alpha}{\alpha}
\int\limits_{-1}^1
\frac{d\beta}{\beta} \theta(\alpha\beta - x_q)\int_{|\alpha| < |A| < 1}
\frac{dA}{A}\int_{|\beta| < |B| < 1} \frac{dB}{B} \theta(AB) =
 \frac{1}{6}
\ln^4(x_q).
\end{equation}
Because the essential momenta $q$ and $Q$ are soft in DL approximation
 we can again neglect the
dependence on them in the numerator in DL approximation. The numerator
of the integrand then takes the form
$$
- \frac{\alpha_sC_F}{4\pi}\frac{4x_q}{4m_qm_H^6} d_{\mu\nu} 
Sp\biggl[ \gamma^
{\mu}\left( {\hat k_1} + {\hat q} + m_q \right)
\gamma^{\lambda}\left(
{\hat k_1} + {\hat Q} + m_q \right)\left( {\hat k_2} 
- {\hat Q} - m_q \right)
\gamma_{\lambda}\times
$$
\begin{equation}\label{18d}
\left( {\hat k_2} - {\hat q} - m_q \right)\gamma^{\nu}\left(
{\hat q} + m_q \right) \biggr] = - 4x_q\frac{\alpha_sC_F}{4\pi}.
\end{equation}
Combining Eqs. (\ref{18c}, \ref{18d}) we obtain
\begin{equation}\label{18e}
F^{(1b)}(x_q) = - \frac{1}{6}\frac{\alpha_sC_F}{4\pi}\ln^2(x_q)
F(x_q)^{(one-loop)}.
\end{equation}

Remaining  two-loop diagrams of Fig. 1(b-e) do not contribute to the
DL asymptotics of the amplitude because both the electromagnetic 
vertex convoluted with a polarization vector of real
external photon and the quark propagator do not contain double logarithms.
These properties can be easily checked at the one loop level.
Only one large logarithm may appear in these structures in the Feynman 
gauge but not the double one.

 Thus, we have restored the exact result (\ref{16}). 
It is clear from (\ref{16})
that when the mass of the Higgs boson is large enough so that in 
spite of smallness of the QCD coupling constant $\alpha_s$
\begin{equation}\label{18}
\frac{\alpha_sC_F}{4\pi}\ln^2(x_q) \sim 1,
\end{equation}
the perturbative QCD expansion in $\alpha_s$ is not valid and we
should modify two-loop result.

In the next section we obtain the quark formfactors $F$ in all
orders of QCD in the following region of Higgs boson masses 
\begin{equation}\label{19}
\frac{\alpha_sC_F}{4\pi} \ll \frac{\alpha_sC_F}{4\pi}\ln(1/x_q) \ll
\frac{\alpha_sC_F}{4\pi}\ln^2(1/x_q) \sim 1.
\end{equation}
In this region we  expect to get the correct 
result in the framework of the DLA.

\section{Resummation and Discussion}

The resummed form factor can be presented in the form
(see Fig. 2)
$$
F(x_q) = \frac{8\pi i}{\left( G_F \sqrt{2} \right)^{1/2} \alpha 
N_c Q_q^2
m_H^4} \int\limits_{-\infty}^{\infty} \frac{d^4q}{(2\pi)^4} d_{\mu\nu}
$$
\begin{equation}\label{21}
\times tr\left[ S(k_1 + q) V(H \rightarrow q{\bar q}) S(q - k_2)
R^{\mu\nu}(q{\bar q} \rightarrow \gamma\gamma) \right],
\end{equation}
where the amplitude $R^{\mu\nu}(q{\bar q} \rightarrow \gamma\gamma)$ 
is defined in such a way that it contains no two-quark cuts 
in the $\gamma\gamma$- channel, $ V(H \rightarrow q{\bar q})$
is the exact QCD vertex which contains 
all possible gluon exchanges, $S$ is the exact quark propagator.
\begin{figure}
\begin{center}
\epsfig{figure=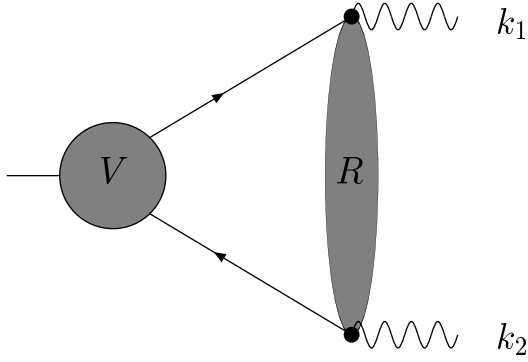}
\end{center}
\caption{The amplitude in all orders of $\alpha_S$.}
\end{figure}
Double logarithms appear when we have consistent logarithmic
integrations. Therefore, only region (\ref{10}) can yield the DL 
contribution to the
RHS of Eq. (\ref{21}) after integration over $q$. 
As it is clear from the above
discussion we should not consider QCD corrections to the amplitude 
$R^{\mu\nu}(q{\bar q} \rightarrow\gamma\gamma)$ and to
the quark propagators $S$ in the DLA, taking them in the Born
approximation. One has to know only the expression for the
$H \rightarrow q{\bar q}$ vertex $V(p_2, p_1)$ in DLA in the following
kinematical region
$$
m_q^2 \ll |p_1^2|, |p_2^2| \ll (p_2 - p_1)^2 = m_H^2,
$$
\begin{equation}\label{23}
p_2 = k_1 + q,\ \ \ p_1 = q - k_2.
\end{equation}
Let us remind that in general $F$ is proportional to $m_q^2$.
One power of $m_q$ appears from the $Hq{\bar q}$- vertex, the 
another one - from the trace over Dirac indexes.
The physical reasons for that were explained in the previous section.
 There was also noted that one of the two powers of $m_q$ 
appears from the quark propagator $({\hat q} + m_q)/(q^2 - m^2_q)$
from the $R^{\mu\nu}(q{\bar q} \rightarrow\gamma\gamma )$ amplitude. 
Due to this circumstance we are not forced to consider
the $O(m_q)$ corrections to the $Hq{\bar q}$- vertex. Therefore, 
the exact result for this vertex in DLA is given
by multiplication of the Born expression with the well-known 
Sudakov form factor
\begin{equation}\label{28}
exp\left[ -\frac{\alpha_sC_F}{2\pi}\ln
\left| \frac{(p_2 - p_1)^2}{p_2^2}
\right| \ln\left| \frac{(p_2 - p_1)^2}{p_1^2} \right| \right].
\end{equation}
Therefore, the quark form factor $F$ in the DLA can be found 
from Eqs. (\ref{13}) and 
(\ref{14})
by multiplying the integrand in the RHS of this
equation with the Sudakov form factor (\ref{28}) and the result reads
\begin{equation}\label{29}
F(x_q) = 4x_q \int\limits_{-1}^1 \int\limits_{-1}^{1} 
\frac{d\alpha}{\alpha} \frac{d\beta}
{\beta} \theta (\alpha\beta - x_q) exp\left[ - 
(\alpha_sC_F/2\pi)\ln|\alpha|
\ln|\beta| \right].
\end{equation}
The integration over $\beta$ can be easily performed and we have
\begin{equation}\label{29a}
F(x_q) = \frac{8x_q\ln^2(x_q)}{Z}\int\limits_{0}^{1}\frac{dy}{y}
\left( 1-exp(-Zy(1-y)) \right) ,
\end{equation}
where $Z = \alpha_sC_F\ln^2(x_q)/(2\pi)$.
We have found that following representation is convenient 
$$
F(x_q) = F(x_q)^{(one-loop)} \sum_{n=0}^{\infty} \frac{2\Gamma(n+1)}
{\Gamma(2n+3)} \left( - \frac{\alpha_sC_F}{2\pi}\ln^2(x_q) \right)^n =
$$
\begin{equation}\label{30}
F(x_q)^{(one-loop)}\left[ 1 - \frac{1}{6}\left( 
\frac{\alpha_sC_F}{4\pi}
\ln^2(x_q) \right) + \frac{1}{45}\left( 
\frac{\alpha_sC_F}{4\pi}\ln^2(x_q)
\right)^2 + ... \right].
\end{equation}
The Eqs. (\ref{29},\ref{30}) are our final results. They correctly 
describe
the quark form factor $F$ in the region of the Higgs boson masses where
the relations (\ref{19}) are fulfilled.

 In Fig.3 we plot relative contributions of two-loop (dashed curve), 
three-loop (dashed-dotted curve), four-loop (dotted curve)
order results (\ref{30}) end exact result in DLA (\ref{29a}) 
(solid curve) 
as a function of $Z$. 
All results are normalized on the one-loop result. 
We see that at small enough $x_q$, large Z, 
the perturbation theory does not work  
and the resummation of DL is mandatory. The middle curve in Fig. 3 
corresponds to the exact in DLA result. We see that the later 
slightly decreases the decay width with increasing $Z$. 
\par 
In conclusion, we have discussed a QCD perturbative 
corrections to the  
$H\to \gamma\gamma$ process in the limit of small quark 
masses in the loops.
We have found a new manifestation of the non-Sudakov type of double 
logarithms and have performed one- and two-loop
calculation to find the origin of such double logarithms. 
Then we have resummed them and have 
derived the exact result for the quark-loop form factor $F$
with double logarithmic accuracy (\ref{29},\ref{30}). 
These results will be important in the situation when the Higss 
boson is heavy enough to guarantee smallness of $m_q/m_H$ and 
when the  mass $ m_q$ itself is not very small to give 
visible contribution to the amplitude. 
Such a situation can be realized 1) in the Standard Model 
with $M_H\approx 500 GeV$ for the $b$ quark loop and 
2) in the Supersymmetric extension of the
Standard Model, where the coupling of $b$ quark 
with Higgs can be large.
For the first application we need to measure the 
decay width of the $H\to \gamma\gamma$ process 
with very high precision. The feasibility 
of that depends on experimental side, of course, and 
should be studied separately. 
We should understand, of course, that at very high Higgs 
masses two-loop 
electroweak corrections became very important due to terms 
proportional to $G_Fm_H^2$. Fortunately, they have been calculated 
\cite{YaMeKo}.  The second application seems to be very promising and 
important phenomenologically.
The crucial fact here will be the value of Higgs mass 
and the value of $tg(\beta)$.

The main purpose of this paper was to attract 
an attention to new non-Sudakov 
type of double logarithms which appears in 
the $H\to\gamma\gamma$ process.

 In our opinion, the results presented here may have 
applications to other cases. 
One of the most interesting is $\gamma\gamma \to q\bar q$ in the
$J_Z=0$ channel, where such non-Sudakov, mass-suppressed DL terms are
phenomenologically important, as it was demonstrated in Ref. [15].

\par    
{\bf Acknowledgements: }
We are grateful to V.S. Fadin, A. Khodjamirian, V.A. Khoze,  
K. Melnikov, R. R\"uckl for useful discussions.
This work is supported by the German Federal Ministry for 
Research and Technology (BMBF) under contract number 05 7WZ91P (0).
  
\begin{figure}[htb]
\epsfxsize=15cm
\centerline{\epsffile{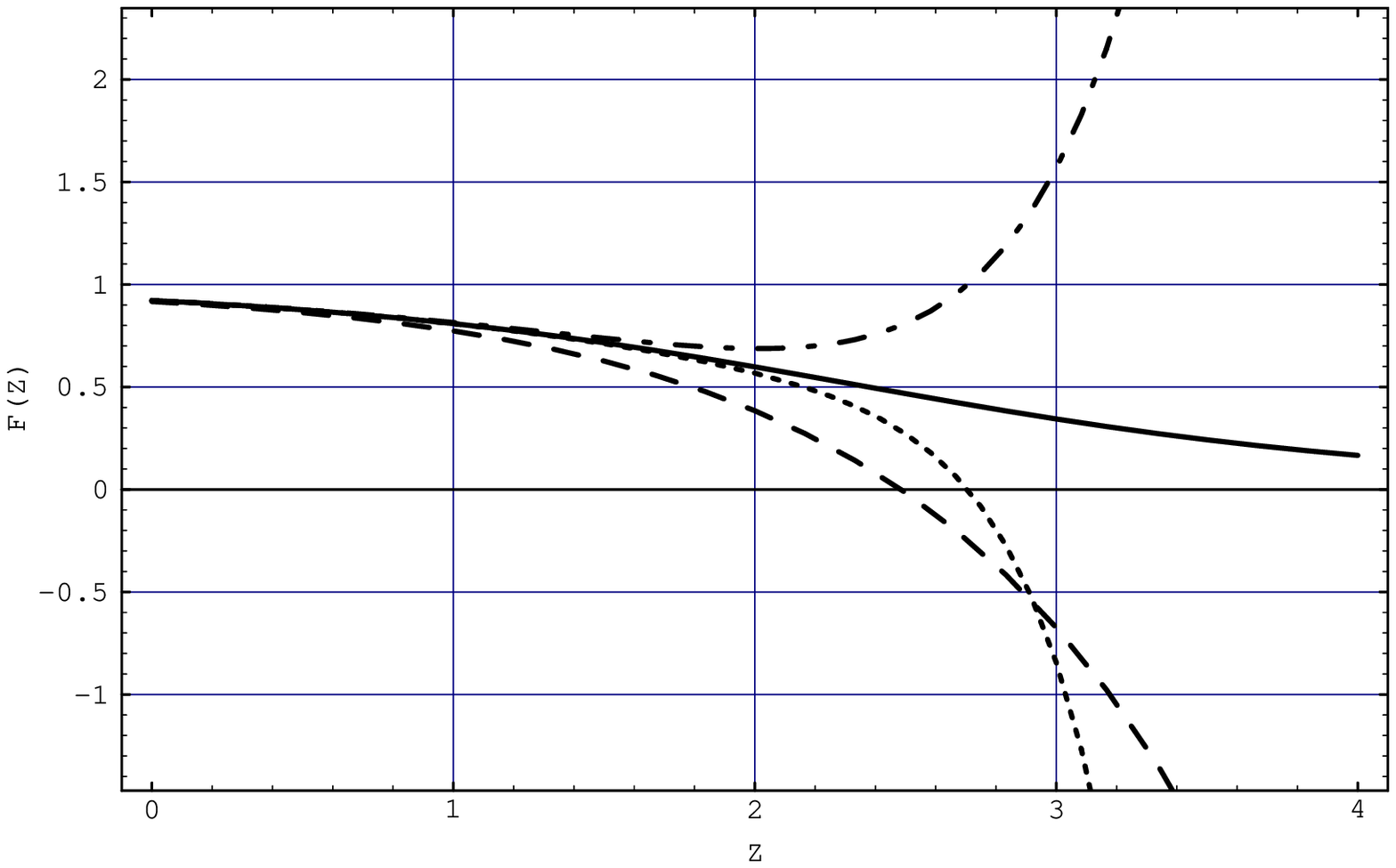}}
\caption[]{The relative contribution of the two-loop (dashed curve), 
the three-loop (dashed-dotted curve), the four-loop (dotted curve) and 
the exact in DLA result for the formfactor (\ref{30}) 
(solid curve) as a function of $Z$. All results are  
normalized on the one-loop result.
The one-loop contribution corresponds to one in such a normalization. }
\end{figure}

\end{document}